\documentclass[doublecol]{epl2} 

\def\be{\begin{equation}}
\def\ee{\end{equation}}
\def\bc{\begin{center}} 
\def\ec{\end{center}}
\def\bea{\begin{eqnarray}}
\def\eea{\end{eqnarray}}
\newcommand{\avg}[1]{\langle{#1}\rangle}

\title{The entropy of randomized  network ensembles  }
\shorttitle{The entropy of randomized  network  ensembles} 

\author{Ginestra Bianconi}
\shortauthor{G. Bianconi}

\institute{The Abdus Salam International Center for Theoretical 
Physics, Strada Costiera 11, 34014 Trieste, Italy            
}

\abstract{Randomized network ensembles are the null models of  real
  networks and are extensively used to compare a real system to a null
  hypothesis. 
In this paper we study network ensembles with the same degree
distribution, the same degree-correlations {and}  the same community
structure of any given  real network. We characterize these randomized network
ensembles by their 
entropy, i.e. the normalized logarithm of the total number of networks
which are part of these ensembles.
 We estimate the entropy of randomized
ensembles starting from a large set of real directed and undirected
networks. We propose entropy as an indicator to  assess
the role of each structural feature in a given  real  network.We observe that the  ensembles with fixed scale-free degree distribution
have smaller entropy than the ensembles with homogeneous  degree
distribution indicating a higher level of order in  scale-free networks.}
\begin{document}

\maketitle
\section{Introduction}
The complexity of a network\cite{BarabasiNewman} depends  on its
global structural organization  which is linked to the functional
constraints the network has to satisfy.
Real networks  show different levels of 
organization.
To characterize  their  structure few different quantities have been
proposed: (i)
the density of the links, (ii) the degree sequence \cite{BA}, (iii)
the degree-degree correlations\cite{Vespignani_corr,Sneppen,Berg}, (iv) the
clustering coefficient \cite{SW,Modular}, (v) the k-core structure \cite{k-core_kirk,k-core_doro,k-core_vesp}
and finally (vi) the community structure \cite{Newman1,Danon,Newman2,review}.
 
To study the different information content retained by these structural
quantites we will consider    randomized network
models which  are best studied by statistical mechanics methods.
Out of different statistical mechanics approaches of networks\cite{Burda_stat,Doro_stat},
one has been  proposed \cite{Park,Caldarelli}  
for networks with hidden variables $\theta_i$ associated to each node 
$i$ of the network.
In the same framework it has been  shown by  \cite{Garlaschelli} 
 that the probability of a link
should satisfy specific forms in order to guarantee good inference of
the hidden variables. 

Every real  network can be considered as  a  specific instance of a
particular network evolution compatible to its functional constraints. 
Nevertheless in many cases real  networks are not determined exactly by their
evolution.
We propose here to consider a real network as   belonging to an
ensemble of networks which would perform the same task equally well. 
For example in the biological world we observe a certain variability
of biological networks across different species with the same
biological function. 
The complexity of a given ensemble of networks increases as the number
of   networks in the ensemble decreases. Consequently a high complexity
of the network ensemble corresponds to a small variability of
the networks in the ensemble. The entropy $\Sigma$ of a given network ensemble \cite{Burda_en} is proportional to the logarithm of the number of
networks  belonging to the ensemble. We expect that a very complex
network is belonging to an ensemble of functionally equivalent networks 
  of small entropy.  
Since it is difficult to characterize the minimal  entropy ensemble a
real network belongs to, we take  successive approximations of the
real network.

To characterize the complexity of a real network we consider  a series of
randomized network models which retain some characteristics of the
real networks. In particular we consider networks with a  given degree
sequence,  given degree-degree correlations and a given  community structure.
Degree-degree  correlation \cite{Berg} has been considered a signature of non randomness
in the topology of the networks. The correlations have been shown to be
important in the Internet at the Autonomous System
Level \cite{Vespignani_corr} and in biological networks \cite{Sneppen}
where the degree correlations are linked also to the modular
structure \cite{Modular} of the network.

In our  approach we will first consider a particular
real network to be  part of the 
ensemble of networks with the same number  of nodes $N$ and links $L$ the real
network has.
This network ensemble is  the $G(N,L)$ studied by the random
graph community.
Subsequently we consider the configuration model of networks with given
degree sequence and we restrict the number of possible networks.
Furthermore we  consider the ensemble of networks with a given degree 
sequence and  with given degree correlations or with given  community structure and we
further restrict the space of possible networks.
Finally we will consider the ensemble of networks with given community
structure and degree sequence. 
How much information is carried by each of these ensembles?
This paper is trying to answer this question  by calculating the 
entropies of these ensembles which  subsequently approximate the  real network.

The ensemble of networks with a given degree sequence falls in the class
of hidden variable models \cite{Park,Caldarelli} with the hidden
variable being nothing else than the Lagrangian multipliers of the
connectivity of each node.

The ensemble of networks with given degree sequence and degree
correlations, or given degree sequence and given community structure
are   generalized hidden variable model and can also be
used to generate  networks with given degree-degree
correlations/community structure.

\section{Undirected networks}
 
Given a real network with $N$ nodes and given adjacency matrix $(a_{ij})$, $i=1,\dots, N$ we construct
subsequent  randomized networks ensembles. 
For an undirected network the first ensemble (zero order
approximation) is the {$G(N,L)$} network ensemble of networks
with given number of nodes $N$ and links $L=\sum_{i,j}a_{ij}/2$. 
The first order approximation is the  configuration network of given
degree sequence $\{k_1,\dots,k_N\}$ with $k_i=\sum_j a_{ij}$. The second order approximation is
the ensemble with given degree sequence $\{k_1,\dots,k_N\}$ and given
average nearest neighbour connectivity
$k_{nn}(k)=\avg{\delta(k_i-k)\sum_j a_ij k_j}$.Moreover one can consider the partition function of the networks with given community structure, and fixed number of links in within each community and between different communities. If the community $q$ of node $i$ is indicated with $q_i$ we can consider graphs with given $A(q,q')=\sum_{i<j}\delta(q_i-q) \delta(q_j-q')a_{ij}$.
The partition functions  of these network ensembles are given by
\bea
Z_0&=&\sum_{\{a_{ij}\}}\delta(L-\sum_{i<j} a_{ij})\exp[\sum_{i<j} h_{ij}a_{ij}]\nonumber \\
Z_1&=&\sum_{\{a_{ij}\}}\prod_i\delta(k_i-\sum_{j} a_{ij})\exp[\sum_{i<j}
h_{ij}a_{ij}]\nonumber \\
Z_2&=&\sum_{\{a_{ij}\}}\prod_i\delta(k_i-\sum_{j} a_{ij})\exp[\sum_{i<j}
h_{ij}a_{ij}]\nonumber \\
& &
\prod_k\delta(k_{nn}(k)N_kk-\sum_{ij}\delta(k_i-k)a_{ij}k_j)\nonumber\\
Z_c&=&\sum_{\{a_{ij}\}}\prod_i\delta(k_i-\sum_{j} a_{ij})\exp[\sum_{i<j}
h_{ij}a_{ij}]\nonumber \\
& &
\prod_{q,q'}\delta(A(q,q')-\sum_{i<j}\delta(q_i-q)\delta(q_j-q')a_{ij})
\eea
where $h_{ij}$ are auxiliary fields, $N_k$ indicates the number of nodes of degree $k$ in the network
$N_k=\sum_i\delta(k_i-k)$, the vector $q_i$ indicates to which
community a node belongs  and $A(q,q')$
indicates the number of links between the community $q$ and  the
community $q'$.
The probability $p_{ij}$ for a link between node $i$ and node $j$ (the probability for $a_{ij}=1$) is given by
\be
p_{ij}^{(\kappa)}=\left.\frac{\partial \ln(Z_{\kappa)}}{\partial h_{ij}}\right|_{h_{ij}=0\forall i,j}
\ee
The number of undirected simple networks  in each of these ensembles $\kappa$
is consequently  given by 
\be
{\cal N}_{\kappa}=\left.Z_{\kappa}\right|_{h_{ij}=0\forall i,j}.
\ee

We define the entropy per node $\Sigma$ of the network ensemble $\kappa$ as
\be
\Sigma_{\kappa}=\frac{1}{N}\ln {\cal N}_{\kappa}.
\ee

The number of undirected networks ${\cal N}_0$ with given number of nodes $N$ and
links $L$ is  given by the binomial
\bea
{\cal N}_0=\left(\begin{array}{c} \frac{N(N-1)}{2}\\ L\end{array}\right)
\eea
for distinguishable nodes in the networks \cite{Burda_en}.
The probability $p_{ij} $of a given link $(i,j)$ is given by
$p_{ij}^{(0)}={L}/({N(N-1)/2})$ for every couple of nodes $i,j$.

\subsection{The volume of the network ensemble with given degree
  sequence}

The first level of approximation is the one in which a given degree
sequence is assumed.
In the undirected simple case the partition function of the network
ensemble  with given
degree distribution is given by 
\be
{Z}_1=\sum_{\{a_{ij}\}}\prod_i\delta(k_i-\sum_j a_{ij})e^{\sum_{i<j}h_{ij}a_{ij}}
\ee 
Expressing the delta's in the integral form with Lagrangian multipliers
$\omega_i$ for every $i=1,\dots N$ we get
\be
{Z}_1=\int {\cal D}\omega \ e^{-\sum_i\omega_i k_i}\prod_{i<j}\left(1+e^{\omega_i+\omega_j+h_{ij}}\right)
\ee
{where ${\cal D}\omega=\prod_i d\omega_i/(2\pi)$.}
 {We solve this integral by saddle point equations}.
The entropy of this ensemble of networks  {can be approximated in the
large network limit with}
 {\bea
N\Sigma_1^{und}&\simeq&-\sum_i\omega_i^*k_i+\sum_{i<j}\ln(1+e^{\omega_i^*
  +\omega_j^*})\nonumber \\
&&-\sum_i \ln(2\pi\alpha_i)
\eea
}
with the Lagrangian multipliers $\omega_i$ satisfying the saddle point equations
\be
k_i=\sum_{j\neq i} \frac{e^{\omega_i^*+\omega_j^*}}{1+e^{\omega_i^*+\omega_j^*}},
\label{sp1}
\ee
 {
and the coefficients $\alpha_i $ defined as
\be
\alpha_i=\sum_j\frac{e^{\omega_i^*+\omega_j^*}}{\left(1+e^{\omega_i^*+\omega_j^*}\right)^2},
\ee
}
The probability of a link $i,j$ in this ensemble is then given by
\be
p_{ij}^{(1)}= \frac{e^{\omega_i^*+\omega_j^*}}{1+e^{\omega_i^*+\omega_j^*}}
\ee
recovering the hidden variable ensemble \cite{Park,Garlaschelli}.
In particular in this ensemble  $p_{ij}\neq f(\omega_i)f(\omega_j)$,
consequently  the model retains  some 'natural' correlations\cite{Garlaschelli}  given by the
degree sequence and the constraint that we consider only simple
networks.
This in fact are nothing else than the  correlations of the
configuration model \cite{Molloy_Reed}.
Nevertheless we can consider  the case in which the network is sparse and
there is a structural cutoff in the system, $k_i<\sqrt{\avg{k}N}$.
In this case we can approximate Eq. $(\ref{sp1})$ by $e^{\omega_i}=k_i
\sqrt{\avg{k}N},\alpha_i=k_i$.
In this limit the network is not correlated
$p_{ij}^{(1),uncorr}=k_ik_j/(\avg{k}N)$,
$\omega^*_i<0$ and we can approximate the entropy of the ensemble as
 {\bea 
N\Sigma_{1,undir}^{uncorr}&\simeq&-\sum_i \ln[k_i/\sqrt{\avg{k}N}]k_i
-\frac{1}{2}\sum_i\log(2\pi k_i)\nonumber \\
&&+\frac{1}{2}\sum_{ij} \frac{k_i
  k_j}{\avg{k}N}-\sum_{ij}\frac{1}{2}\frac{k_i^2 k_j^2}{(\avg{k}N)^2}+\dots\nonumber \\
&= &-\sum_i (\ln k_i-1) k_i-\frac{1}{2}\sum_i \ln(2\pi k_i)+
\nonumber \\
&& \frac{1}{2}\avg{k}N
[\ln(\avg{k}N)-1] -\frac{1}{2}\left(\frac{\avg{k^2}}{\avg{k}}\right)^2+\dots
\eea
}
which approximately gives for the volume
\be
{\cal N}_1^{uncorr}\simeq\frac{(\avg{k}N)!!}{\prod_i
  k_i!}\exp\left[-\frac{1}{2}\left(\frac{\avg{k^2}}{\avg{k}}\right)^2\right].
\label{N_1_unc}
\ee 
The expression $(\ref{N_1_unc})$ was already derived in \cite{Chaos} by
combinatorial considerations valid in a network with structural cutoff.
In fact the term $(\avg{k}N)!!\simeq (\avg{k}N-1)!!$ gives the total number of different ways we can
link the $2L=\avg{k}N$ half-edges associated to  a degree sequence to form a network.
In fact we can take a first half-edge of the network and we have $2L-1$
choices to match it with one of the other half edges. then we can take
another half-edges and we have $(2L-3)$ possible choices of other
half-edges to link to, giving rise to $(2L-1)!!$ networks.
Out of these networks only a part of them is simple providing for the
correction
$\exp\left[-\frac{1}{2}\left(\frac{\avg{k^2}}{\avg{k}}\right)^2\right]$\cite{Chaos}.
Out  of  these simple networks for each  distinct adjacency matrix
there are $\prod_i k_i!$ networks that can be constructed by simply
permuting the order of the edges at each node.

\begin{figure}
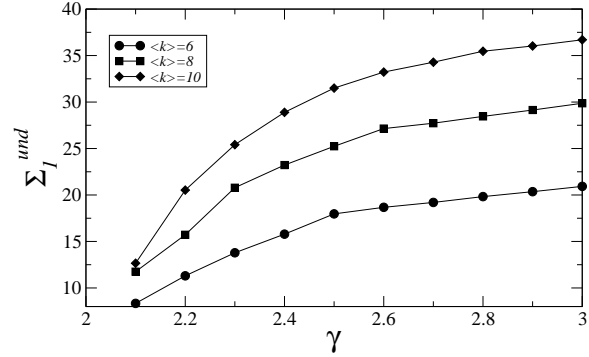

\onefigure{fig1.eps}
\caption{The entropy $\Sigma_1^{und}$
 of a configuration model of a network with  power-law degree
 distribution $N=10^4$ nodes and a fixed average connectivity
 $\avg{k}=6,8,10$ as a function of the power-law exponent $\gamma$.}
\label{SF.fig}
\end{figure}

It can be shown that  within sparse uncorrelated networks the
scale-free networks with $\gamma\rightarrow 2$ are the ones 
which minimize ${\Sigma}_1^{unc}$
\cite{Chaos}.
For correlated networks with natural correlations the entropy of the
configuration model $\Sigma_1$ decreases with the value of the
power-law exponent $\gamma$.
In figure \ref{SF.fig} we plot the entropy of a scale-free network
with natural cutoff and fixed average connectivity $\avg{k}=6,8,10$.
The entropy $\Sigma_1$ of the configuration model  is decreasing with
decreasing  power-law exponent $\gamma$ reaching its minimum at
$\gamma\rightarrow 2$.
This indicates that scale-free networks with low value of $\gamma$
presents higher level of ordering with respect to random homogeneous
networks. 

\subsection{ The volume of a network ensemble with fixed degree correlations}
The second order of approximation is to take into consideration degree
correlations behind the 'natural correlations' of the configuration
model.
The partition function for this ensemble is given by
\bea
Z_2&=&\sum_{\{a_{ij}\}}\prod_i\delta(k_i-\sum_j
a_{ij})e^{\sum_{i<j}h_{ij}a_{ij}}\nonumber \\
& &\prod_{k=1}^K\delta\left[k_{nn}(k)k
  N_k-\sum_{ij} k_j a_{i,j}\delta(k_i-k)\right]
\eea 
{where $K$ is the maximal connectivity in the network.}
Expressing the deltas in the integral form we get for the partition function
\bea
Z_2&=&\int {\cal D}\omega \int {\cal D}A e^{-\sum_i\omega_i k_i-\sum_k
  A_k  k_{nn}(k)k}\nonumber \\
& &\prod_{i<j}\left(1+e^{\omega_i+\omega_j +h_{ij}+k_j A_{k_i}+k_i A_{k_j}}\right),
\eea
The expression which can be evaluated as for the case of the
camculation of $\Sigma_1$ where the
Lagrange multipliers $\omega_i$ and $A_k$ satisfy 
\bea
k_i&=&\sum_{j\neq i}
\frac{e^{\omega_i^*+\omega_j^*+k_j A_{k_i}^*+k_i
    A_{k_j}}}{1+e^{\omega_i^*+\omega_j^*+k_j A_{k_i}^*+k_i A_{k_j}^*}},
    \\
k_{nn}(k)&=&\frac{1} {k N_k}\sum_i\delta(k_i-k)\sum_{j\neq i} k_j \frac{e^{\omega_i^*+\omega_j^*+k_j A_{k_i}^*+k_i A_{k_j}^*}}{1+e^{\omega_i^*+\omega_j^*+k_j A_{k_i}^*+k_i A_{k_j}^*}}. \nonumber
\label{sp}
\eea
If we solve this equation for a given real network degree sequence and
nearest neighbor average degree, we can then construct other networks
in the same ensemble just by drawing a link $i,j$ with probability
\be
p_{ij}^{(2)}=\frac{e^{\omega_i^*+\omega_j^*+k_j A_{k_i}^*+k_i
    A_{k_j}}}{1+e^{\omega_i^*+\omega_j^*+k_j A_{k_i}^*+k_i A_{k_j}^*}}.
\ee
The entropy of this ensemble is  {approximatly equal in the large network limit to 
\bea
N\Sigma_2^{und}&\simeq&-\sum_i\omega_i^*k_i-\sum_k A_k k_{nn}(k) k N_k\nonumber \\
& &+\sum_{i<j}\ln(1+e^{\omega_i^* +\omega_j^*+k_i A_{k_{j}}+k_j
  A_{k_{i}}})\nonumber \\
&&-\frac{1}{2}\sum_i \ln(2\pi \alpha_i)-\frac{1}{2}\sum_k \ln(2\pi \alpha_k)
\eea

with $\alpha_i,\alpha_k$  defined as
\bea
\alpha_i&=&\sum_j\frac{e^{\omega_i^*+\omega_j^*+k_j A_{k_i}^*+k_i
    A_{k_j}}}{\left(1+e^{\omega_i^*+\omega_j^*+k_j A_{k_i}^*+k_i
    A_{k_j}^*}\right)^2} \\
\alpha_k&=&\sum_i\delta(k_i-k)\sum_{j\neq i} k_j^2 \frac{e^{\omega_i^*+\omega_j^*+k_j A_{k_i}^*+k_i A_{k_j}^*}}{\left(1+e^{\omega_i^*+\omega_j^*+k_j A_{k_i}^*+k_i A_{k_j}^*}\right)^2}.\nonumber
\eea
}
In table \ref{table-und} we report the entropy  for
different undirected network\cite{nnets} ensembles at different level of approximation.
We consider the Internet network at the Autonomous System Level, the Protein Interaction networks
of {\it S. cerevisiae} (DIP database) and  the partial map
of protein interaction network of {\it H. Sapiens} \cite{Vidal}.
We observe that for these networks taking into account the degree
distribution strongly reduces the entropy of the randomized network
ensemble. {Moreover  taking explicitly into account 
for degree-degree correlation fine tune the value of the  entropy of the
randomized ensemble.} 

\subsection{The volume of network ensemble with given community structure}

A different  ensemble of networks is the ensemble of networks 
 with given community structure and degree sequence. 
Suppose that we have a network and we detect   $Q$ communities 
 such that each node $i=1,\dots, N$ belongs to the community
$q_i=1,\dots,Q$ with $Q$ finite.
To find a randomized ensemble of networks with the given community
structure we impose that the nodes have fixed degree sequence  and fixed number 
$A(q,q')$ of links in between the communities $q$ and $q'$.
In an undirected   network,  $A(q,q')$ is given by the following expression
\be
A(q,q')=\sum_{i<j}\delta(q_i-q)\delta(q_j-q')a_{ij}.
\ee  
Following the same steps as in the previous case we find that the
entropy for such an ensemble is given by
 {
\bea
N\Sigma_c&\simeq&-\sum_ik_i\omega_i-\sum_{q\leq q'}A(q,q')w_{q,q'}\nonumber
\\
&&+\sum_{i<j}\ln\left(1+e^{\omega_i+\omega_j+w_{q_i,q'_j}}\right)\nonumber\\
&& -\frac{1}{2}\sum_i \ln(2\pi \alpha_i)-\frac{1}{2}\sum_{q<q'}\ln(2\pi\alpha_{q,q'})
\eea
}
with the Lagrangian multipliers $\{\omega_i\},\{w_{q,q'}\}$ satisfying the saddle point equations
\bea
k_i&=&\sum_{i<j}\frac{e^{\omega_i+\omega_j+w_{q_i,q_j}}}{1+e^{\omega_i+\omega_j+w_{q_i,q_j}}} \\
A(q,q')&=&\sum_{i<j}\delta({q_i-q})\delta(q_j-q')\frac{e^{\omega_i+\omega_j+w_{q,q'}}}{1+e^{\omega_i+\omega_j+w_{q,q'}}} ,\nonumber
\eea 
 {and $\alpha_i,\alpha_{q,q'}$ defined as
\bea
\alpha_i&=&\sum_j\frac{e^{\omega_i+\omega_j+w_{q_i,q_j}}}{\left(e^{\omega_i+\omega_j+w_{q_i,q_j}}\right)}\nonumber\\
\alpha_{q,q'}&=&\sum_{i,j}\delta({q_i-q})\delta(q_j-q')\frac{e^{\omega_i+\omega_j+w_{q,q'}}}{\left(1+e^{\omega_i+\omega_j+w_{q,q'}}\right)^2}
\eea
}
The probability for a link between node $i$ and $j$ is equal to
\be
p_{ij}^{(c)}=\frac{e^{\omega_i+\omega_j+w_{q_i,q_j}}}{1+e^{\omega_i+\omega_j+w_{q_i,q_j}}}.
\ee
In the case of the Zachary club \cite{Zachary} we where able to calculate
 {$\Sigma_1^{undir}=3.94$ and $\Sigma_c^{undir}=3.25$} quantifying the
amount of information present in the known community partition.
\begin{table}
\begin{center}
\begin{tabular}{ l c c c c c }
Network &N & L &  $\Sigma_0^{und}$ &$\Sigma_1^{und}$ &$\Sigma_2^{und}$  \\
AS-97-11 & 3015 & 5156 & 13.3 &7.5 & 7.3  \\
AS-98-10&4180&7768 &14.9 & 8.6 & 8.4   \\
AS-99-10 &5861  &11312 &16.1 &9.2&9.0 \\
AS-00-10 &8836&17822 &17.5& 9.8&  9.6 \\
AS-01-03 &10515 &21455 & 18.1&10.1 & 9.8 \\
Yeast DIP   &4135& 8099&15.6 &12.3 &  11.1 \\
H. Sapiens PI & 3134& 6726&16.3 &12.3 &12.2 \\
\end{tabular}
\end{center}
\caption{Entropies of randomized  network ensembles starting from real
  undirected networks with $N$ nodes and $L$
  links. $\Sigma_0^{und},\Sigma_1^{und},\Sigma_2^{und}$ indicate
  the entropy of a undirected network with assigned $N$ nodes and $L$
  links, with given degree sequence and with given degree sequence and
  degree correlations respectively. The data sets\cite{nnets} ``AS-year-month''
  indicate different snapshot of the Internet at the Autonomous
  System level, the yeast DIP dataset is the protein interaction of
  {\it S. cerevisiae} and {\it H. Sapiens} PI is the partial human protein interaction map \cite{Vidal}. 
\label{table-und}}
\end{table}

\section{ Directed networks}
An undirected network is determined by a symmetric adjacency
matrix, while the matrix of a directed  network is in general
non-symmetric.
Consequently the degrees of freedom of a directed network are more
  than the degrees of freedom of an undirected network.
If we consider the number of directed networks ${\cal N}_0^{dir}$ with
given number of nodes and of directed links we find
\bea
{\cal N}_0^{dir}=\left(\begin{array}{c} N(N-1) \\ L^{dir}\end{array}\right).
\eea
\subsection{Volume of randomized directed network ensembles with given degree sequence}
To calculate the volume of directed networks with a given  degree sequence of
in/out degrees $\{k_i^{out},k_i^{in}\}$ we just have to impose 
the constraints on the incoming and outgoing connectivities,
\bea
Z_1^{dir}&=&\sum_{\{a_{ij}\}}\prod_i\delta(k_i^{(out)}-\sum_j a_{ij})
\prod_i \delta(k_i^{(in)}-\sum_j a_{ji})\nonumber \\
& &\exp[\sum_{ij}h_{i,j}a_{ij}]
\eea 
Following the same approach as for the undirected case,  { we
  find that the entropy  of this ensemble of networks is given by 
\bea
N\Sigma_1^{dir}&\simeq&-\sum_i\omega_i^* k_i^{(out)}-\sum_i k_i^{(in)}
\hat{\omega}_i^*\nonumber \\
& & +\sum_{i\neq j}\ln(1+e^{\omega_i^*+ \hat{\omega}_j^*})\nonumber \\
&&-\frac{1}{2}\sum_i \ln((2\pi)^2 \alpha^{(in)}_i \alpha^{(out)}_i)
\eea
}
with the Lagrangian multipliers satisfying the saddle point equations
\bea
k_i^{(out)}&=&\sum_{j\neq i}
\frac{e^{\omega_i^*+\hat{\omega}_j^*}}{1+e^{\omega_i^*+\hat{\omega}_j^*}}.\nonumber
\\
k_i^{(in)}&=&\sum_{j\neq i}
\frac{e^{\omega_j^*+\hat{\omega}_i^*}}{1+e^{\omega_j^*+\hat{\omega}_i^*}}.
\label{sp}
\eea
 {
with 
\bea\alpha^{(out)}_i&=&\sum_{j\neq i}
\frac{e^{\omega_i^*+\hat{\omega}_j^*}}{(1+e^{\omega_i^*+\hat{\omega}_j^*})^2}\nonumber \\
\alpha_i^{(in)}&=&\sum_{j\neq i}
\frac{e^{\omega_j^*+\hat{\omega}_i^*}}{(1+e^{\omega_j^*+\hat{\omega}_i^*})^2}
\eea
}
The probability for a directed link from $i$ to $j$ is given by
\be
p_{ij}^{(1,dir)}=\frac{e^{\omega_i^*+\hat{\omega}_j^*}}{1+e^{\omega_i^*+\hat{\omega}_j^*}}.
\ee
 {
\begin{table*}
\begin{center}
\begin{tabular}{ l c c c c c c }
~~Network &N & L &  $\Sigma_0^{dir}$& $\Sigma_0^{und}$&$\Sigma_1^{dir}$ &$\Sigma_1^{und}$  \\
Littlerock FW & 183 & 2,494 & 48.4 &38.4 & 13.28  & 23.44  \\
Seagrass FW&48&226& 15.3 & 11.8 &4.3  & 7.8    \\
Metabolic net. & 896 & 964 &8.3 & 7.5 & 3.2 & 4.3\\
Neural net. &306& 2,359& 35.9&30.52 & 17.8 & 22.5 \\
Power-grid net. & 4,888&5,855 &11.1&10.3&7.5 &8.7\\
ND WWW      &325,729& 1,497,135&55.9 &52.7  &33.1 &36.7\\
\end{tabular}
\end{center}
\caption{Entropy of randomized network ensemble starting form specific
  directed networks with
  of $N$ nodes and $L$ links.
 $\Sigma_0^{(dir/undir)}$ is the entropy of the network ensembles with
 fixed number of nodes $N$ and links $L$, in the case of a directed
 ensemble or in the case of an undirected ensemble.
 $\Sigma_1^{(dir/undir)}$ is the entropy of the directed/undirected
 network ensemble with given degree sequence. The datasets\cite{nnets}
 indicate different foodwebs (FW), the metabolic network of {\it
   E.coli}, the Texas power-grid, the Notre Dame University domain
 WWW, the neural network of {\it C.elegans}. \label{table-dir}}
\end{table*}
}
If the $\omega_i+\hat{\omega}_j<0 \forall i,j=1,\dots N$ the directed
network becomes uncorrelated and we have
$p_{ij}^{1,(dir)}=k_i^{(out)}k_j^{(in)}/\sqrt{\avg{k_{in}}N}$.
Given this solution the condition for having uncorrelated directed
networks is  that the maximal in-degree $K^{(in)}$ and the maximal
out-degree $K^{(out)}$ should satisfy,
$K^{(in)}K^{(out)}/\sqrt{\avg{k_{in}}N}<1$.
The entropy of the directed uncorrelated network is then given by 
 {
\bea
N\Sigma_{1,dir}^{uncorr}&\simeq& \ln(\avg{k_{in}}N)!-\sum_i \ln(k_i^{(in)}!k_i^{(out)}!)\nonumber \\
&&-\frac{1}{2}\frac{\avg{k_{in}^2}}{\avg{k_{in}}}\frac{\avg{k_{out}^2}}{\avg{k_{out}}}
\eea 
}
which has a   clear combinatorial
interpretation as it happens also for the undirected case.
In table $\ref{table-dir}$ we report the entropy of directed networks
and their undirected version observing that different degree
distributions reduce the entropy of randomized network ensembles by a different amount,
some carrying more information than others.
\section{Conclusions}
In conclusion we have studied the space of possible networks in randomized
models of complex networks. 
We  have found that random scale-free network ensembles  with low power-law exponent $\gamma$
have a lower entropy than random network with an homogeneous degree
distribution.
The successive random approximations  of a real graph
characterize to which extent the degree sequence, the degree-degree
correlations or  the community structure constraint the network.  
We have evaluated the entropy of randomized ensembles starting from a
set of different real directed and undirected  networks  showing how much
each structure feature reduce the space of possible networks.
 {Future
work will focus in extending these results to weighted networkand
measurement of large deviations in ensembles of random networks with
hidden variables.}
\acknowledgments
This was was supported by the IST STREP GENNETEC contract number
034952, the author acknowledge D. Garlaschelli and M. Marsili for interesting discussions.

\end{document}